\documentclass[runningheads]{llncs}

\usepackage{graphicx}
\usepackage{hyperref}
\usepackage{textcomp}
\usepackage{bbding}
\usepackage[font=scriptsize,labelfont=bf]{caption}
\usepackage{xcolor}
\usepackage{pgf}
\usepackage{tikz}
\usetikzlibrary{arrows, automata, shapes, petri, positioning, calc}
\usepackage[export]{adjustbox}

\hypersetup{
	colorlinks=true,
	linkcolor={red!50!black},
	citecolor={blue!60!black},
	urlcolor={blue!80!black},
	pdfauthor={Pegoraro Marco, Shankara Narayana Madhavi Bangalore, Benevento Elisabetta, van der Aalst Wil M.P., Martin Lukas, Marx Gernot},
	pdftitle={{Analyzing Medical Data with Process Mining: a COVID-19 Case Study}},
	pdfsubject={Process Mining in Healthcare},
	pdfkeywords={Process Mining, Healthcare, COVID-19},
	pdfproducer={LaTeX},
	pdfcreator={pdfLaTeX},
	bookmarksopen=true
}

\begin{document}

\title{Analyzing Medical Data with Process Mining:\\ a COVID-19 Case Study\thanks{
		\mbox{We acknowledge the ICU4COVID project (funded by European Union's Horizon 2020} under grant agreement n. 101016000) and the COVAS project for our research interactions. Please cite as: Pegoraro, Marco, Madhavi Bangalore Shankara Narayana, Elisabetta Benevento, Wil M. P. van der Aalst, Lukas Martin, and Gernot Marx. “Analyzing Medical Data with Process Mining: a COVID-19 Case Study”. In: Workshop on Applications of Knowledge-Based Technologies in Business, AKTB 2021, Hannover, Germany, June 14, 2021. Springer, 2021
}}
\subtitle{Postprint version - Accepted at the Workshop on Applications of Knowledge-Based Technologies in Business (AKTB 2021)}

\author{Marco Pegoraro\inst{1}\Envelope\orcidID{0000-0002-8997-7517} \and Madhavi Bangalore Shankara Narayana\inst{1}\orcidID{0000-0002-7030-1701} \and Elisabetta Benevento\inst{1, 3}\orcidID{0000-0002-3999-8977} \and \\ Wil M.P. van der Aalst\inst{1}\orcidID{0000-0002-0955-6940} \and Lukas Martin\inst{2}\orcidID{0000-0001-8650-5090} \and Gernot Marx\inst{2}}

\authorrunning{Pegoraro et al.}

\institute{Chair of Process and Data Science (PADS) \\ Department of Computer Science, RWTH Aachen University, Aachen, Germany
	\email{\{pegoraro, madhavi.shankar, benevento, vwdaalst\}@pads.rwth-aachen.de}\\
	\and
	Department of Intensive Care and Intermediate Care \\ RWTH Aachen University Hospital, Aachen, Germany
	\email{\\\{lmartin, gmarx\}@ukaachen.de}
	\and
	Department of Energy, Systems, Territory and Construction Engineering, University of Pisa, Pisa, Italy
}

\maketitle

\begin{abstract}
The recent increase in the availability of medical data, possible through automation and digitization of medical equipment, has enabled more accurate and complete analysis on patients' medical data through many branches of data science. In particular, medical records that include timestamps showing the history of a patient have enabled the representation of medical information as sequences of events, effectively allowing to perform process mining analyses. In this paper, we will present some preliminary findings obtained with established process mining techniques in regard of the medical data of patients of the Uniklinik Aachen hospital affected by the recent epidemic of COVID-19. We show that process mining techniques are able to reconstruct a model of the ICU treatments for COVID patients.

\end{abstract}

\section{Introduction}

The widespread adoption of Hospital Information Systems (HISs) and Electronic Health Records (EHRs), together with the recent Information Technology (IT) advancements, including e.g. cloud platforms, smart technologies, and wearable sensors, are allowing hospitals to measure and record an ever-growing volume and variety of patient- and process-related data~\cite{koufi2015big}. This trend is making the most innovative and advanced data-driven techniques more applicable to process analysis and improvement of healthcare organizations~\cite{galetsi2020review}.
Particularly, \emph{process mining} has emerged as a suitable approach to analyze, discover, improve and manage real-life and complex processes, by extracting knowledge from event logs~\cite{van2016data}. Indeed, healthcare processes are recognized to be complex, flexible, multidisciplinary and ad-hoc, and, thus, they are difficult to manage and analyze with traditional model-driven techniques~\cite{mans2015process}. Process mining is widely used to devise insightful models describing the flow from different perspectives---e.g., control-flow, data, performance, and organizational.

On the grounds of being both highly contagious and deadly, COVID-19 has been the subject of intense research efforts of a large part of the international research community. Data scientists have partaken in this scientific work, and a great number of articles have now been published on the analysis of medical and logistic information related to COVID-19. In terms of raw data, numerous openly accessible datasets exist. Efforts are ongoing to catalog and unify such datasets~\cite{guidotti2020covid}. A wealth of approaches based on data analytics are now available for descriptive, predictive, and prescriptive analytics, in regard to objectives such as measuring effectiveness of early response~\cite{lavezzo2020suppression}, inferring the speed and extent of infections~\cite{anastassopoulou2020data,sarkar2020modeling}, and predicting diagnosis and prognosis~\cite{wynants2020prediction}. However, the process perspective of datasets related to the COVID-19 pandemic has, thus far, received little attention from the scientific community.

The aim of this work-in-progress paper is to exploit process mining techniques to model and analyze the care process for COVID-19 patients, treated at the Intensive Care Unit (ICU) ward of the Uniklinik Aachen hospital in Germany. In doing so, we use a real-life dataset, extracted from the ICU information system. More in detail, we discover the patient-flows for COVID-19 patients, we extract useful insights into resource consumption, we compare the process models based on data from the two COVID waves, and we analyze their performance. The analysis was carried out with the collaboration of the ICU medical staff.

The remainder of the paper is structured as follows. Section~\ref{sec:dataset} describes the COVID-19 event log subject of our analysis. Section~\ref{sec:analysis} reports insights from preliminary process mining analysis results. Lastly, Section~\ref{sec:conclusion} concludes the paper and describes our roadmap for future work.

\section{Dataset Description}\label{sec:dataset}

\begin{figure}[t]
\centering
\includegraphics[width=.7\textwidth]{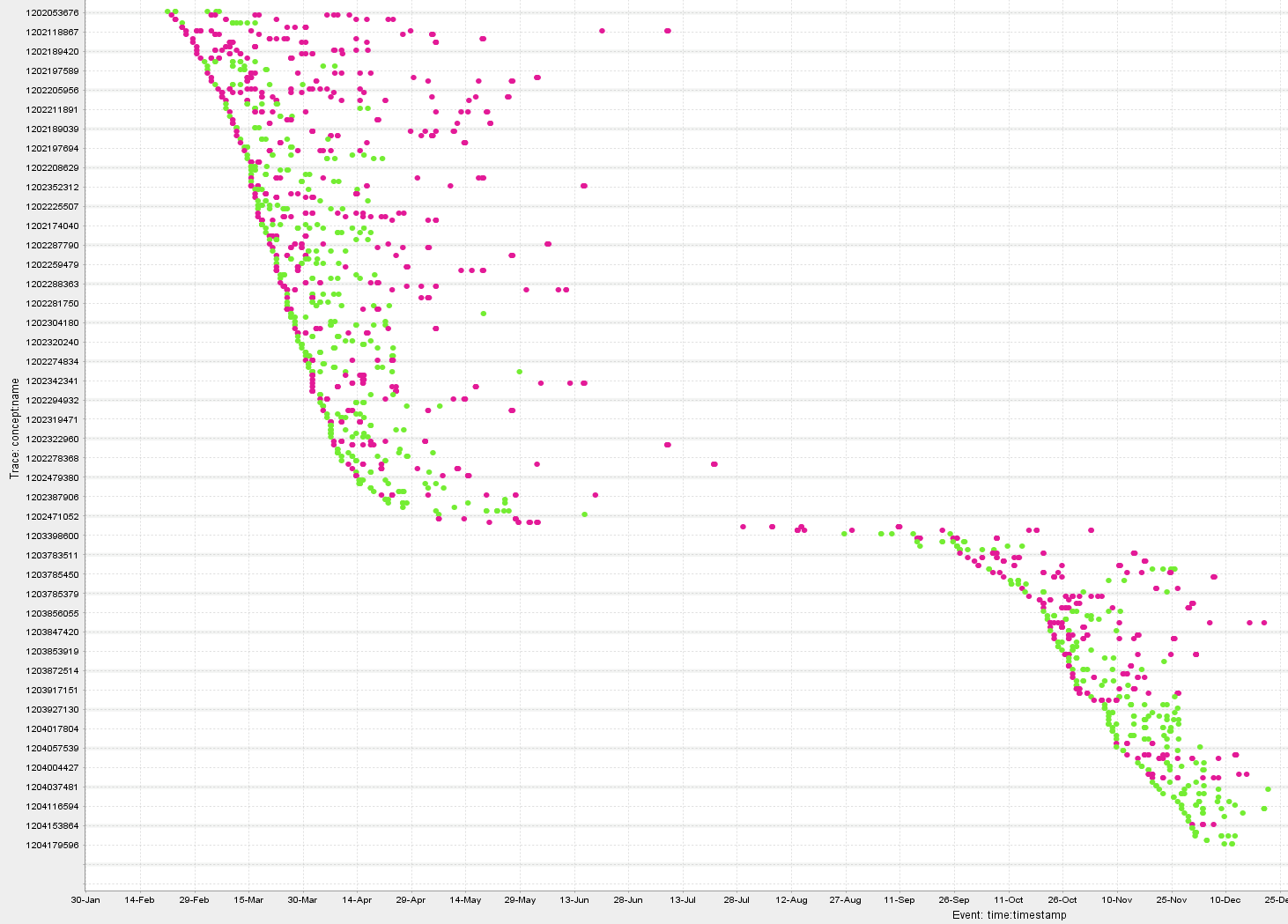}
\caption{Dotted chart of the COVAS event log. Every dot corresponds to an event recorded in the log; the cases with Acute Respiratory Distress Syndrom (ARDS) are colored in pink, while cases with no ARDS are colored in green. The two ``waves'' of the virus are clearly distinguishable.}
\label{fig:dotted_chart}
\end{figure}

The dataset subject of our study records information about COVID-19 patients monitored in the context of the COVID-19 Aachen Study (COVAS). The log contains event information regarding COVID-19 patients admitted to the Uniklinik Aachen hospital between February 2020 and December 2020. The dataset includes 216 cases, of which 196 are complete cases (for which the patient has been discharged either dead or alive) and 20 ongoing cases (partial process traces) under treatment in the COVID unit at the time of exporting the data. The dataset records 1645 events in total, resulting in an average of 7.6 events recorded per each admission. The cases recorded in the log belong to 65 different variants, with distinct event flows. The events are labeled with the executed activity; the log includes 14 distinct activities. Figure~\ref{fig:dotted_chart} shows a dotted chart of the event log.

\section{Analysis}\label{sec:analysis}
In this section, we illustrate the preliminary results obtained through a detailed process mining-based analysis of the COVAS dataset. More specifically, we elaborate on results based on control-flow and performance perspectives.

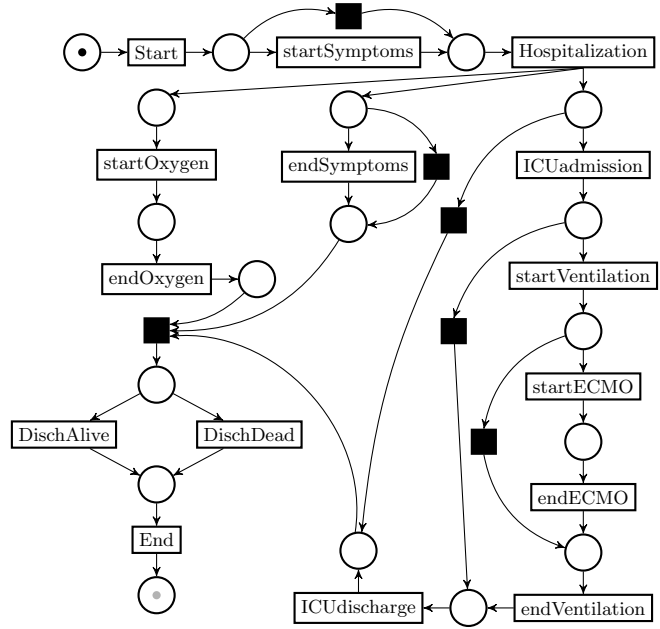
\begin{figure}[t]
	\centering
	\begin{adjustbox}{center}
		\begin{tikzpicture}[node distance=.3cm and .35cm, >=stealth', nodes={scale=.8}]
		
		\tikzstyle{place} = [circle,draw,thick,minimum size=6mm]
		\tikzstyle{transition} = [rectangle,draw,thick,minimum size=4mm]
		\tikzstyle{invisible} = [transition, fill=black]
		\tikzstyle{finaltoken} = [token, fill=black!30]
		
		
		\node [place,tokens=1] (p1) [] {};
		
		\node [transition] (Start) [right= of p1] {Start};
		\draw [->] (p1) to (Start.west);
		
		\node [place] (p2) [right= of Start] {};
		\draw [->] (Start.east) to (p2);
		
		\node [transition] (startSymptoms) [right= of p2] {startSymptoms};
		\node [invisible] (t0) [above= .1 cm of startSymptoms] {};
		\draw [->] (p2) to (startSymptoms.west);
		\draw [->, bend left] (p2) to (t0);
		
		\node [place] (p3) [right= of startSymptoms] {};
		\draw [->] (startSymptoms.east) to (p3);
		\draw [->, bend left] (t0) to (p3);
		
		\node [transition] (Hospitalization) [right= of p3] {Hospitalization};
		\draw [->] (p3) to (Hospitalization.west);
		
		
		\node [place] (p4) [below= of Start] {};
		\node [place] (p5) [below= of startSymptoms] {};
		\node [place] (p6) [below= of Hospitalization] {};
		\draw [->] (Hospitalization.south) to (p4.45);
		\draw [->] (Hospitalization.south) to (p5.45);
		\draw [->] (Hospitalization) to (p6);
		
		\node [transition] (startOxygen) [below= of p4] {startOxygen};
		\draw [->] (p4) to (startOxygen);
		
		\node [place] (p7) [below= of startOxygen] {};
		\draw [->] (startOxygen) to (p7);
		
		\node [transition] (endOxygen) [below= of p7] {endOxygen};
		\draw [->] (p7) to (endOxygen);
		
		\node [place] (p8) [right= of endOxygen] {};
		\draw [->] (endOxygen) to (p8);
		
		\node [transition] (endSymptoms) [below= of p5] {endSymptoms};
		\node [invisible] (t1) [right= .1cm of endSymptoms] {};
		\draw [->] (p5) to (endSymptoms);
		\draw [->, bend left] (p5) to (t1.north);
		
		\node [place] (p9) [below= of endSymptoms] {};
		\draw [->] (endSymptoms) to (p9);
		\draw [->, bend left] (t1.south) to (p9);
		
		\node [transition] (ICUadmission) [below= of p6] {ICUadmission};
		\draw [->] (p6) to (ICUadmission);
		
		\node [place] (p11) [below= of ICUadmission] {};
		\draw [->] (ICUadmission) to (p11);
		
		\node [transition] (startVentilation) [below= of p11] {startVentilation};
		\draw [->] (p11) to (startVentilation);
		
		\node [place] (p12) [below= of startVentilation] {};
		\draw [->] (startVentilation) to (p12);
		
		\node [transition] (startECMO) [below= of p12] {startECMO};
		\draw [->] (p12) to (startECMO);
		
		\node [place] (p13) [below= of startECMO] {};
		\draw [->] (startECMO) to (p13);
		
		\node [transition] (endECMO) [below= of p13] {endECMO};
		\draw [->] (p13) to (endECMO);
		
		\node [place] (p14) [below= of endECMO] {};
		\draw [->] (endECMO) to (p14);
		
		\node [transition] (endVentilation) [below= of p14] {endVentilation};
		\draw [->] (p14) to (endVentilation);
		
		\node [place] (p15) [left= of endVentilation] {};
		\draw [->] (endVentilation) to (p15);
		
		\node [transition] (ICUdischarge) [left= of p15] {ICUdischarge};
		\draw [->] (p15) to (ICUdischarge);
		
		\node [place] (p16) [above= of ICUdischarge] {};
		\draw [->] (ICUdischarge) to (p16);
		
		\node [invisible] (t2) [left= 1.3cm of p12] {};
		\draw [->, bend right] (p11) to (t2.north);
		\draw [->] (t2) to (p15);
		
		\node [invisible] (t3) [left= .9cm of p13] {};
		\draw [->, bend right] (p12) to (t3.north);
		\draw [->, bend right] (t3.south) to (p14);
		
		\node [invisible] (t4) [left= 1.3cm of p11] {};
		\draw [->, bend right] (p6) to (t4);
		\draw [->, bend right=10] (t4) to (p16.80);
		
		
		\node [invisible] (t5) [below= of endOxygen] {};
		\draw [->, bend left=20] (p8) to (t5.25);
		\draw [->, bend left] (p9) to (t5);
		\draw [->, bend right=52] (p16.100) to (t5.-25);
		
		\node [place] (p17) [below= of t5] {};
		\draw [->] (t5) to (p17);
		
		\node [transition] (DischDead) [below right= of p17] {DischDead};
		\draw [->] (p17) to (DischDead);
		
		\node [transition] (DischAlive) [below left= of p17] {DischAlive};
		\draw [->] (p17) to (DischAlive);
		
		\node [place] (p18) [below left= of DischDead] {};
		\draw [->] (DischDead) to (p18);
		\draw [->] (DischAlive) to (p18);
		
		\node [transition] (End) [below= of p18] {End};
		\draw [->] (p18) to (End);
		
		\node [place] (p19) [below= of End] {};
		\draw [->] (End) to (p19);
		\node[finaltoken] at (p19) {};
		
		\end{tikzpicture}
	\end{adjustbox}
	\caption{A normative Petri net that models the process related to the COVAS data.}
	\label{fig:procmodel}
\end{figure}

Firstly, we present a process model extracted from the event data of the COVAS event log. Among several process discovery algorithms in literature~\cite{van2016data}, we applied the Interactive Process Discovery (IPD) technique~\cite{dixit2018interactive} to extract the patient-flows for COVAS patients, obtaining a model in the form of a Petri net (Figure~\ref{fig:procmodel}). IPD allows to incorporate domain knowledge into the discovery of process models, leading to improved and more trustworthy process models. This approach is particularly useful in healthcare contexts, where physicians have a tacit domain knowledge, which is difficult to elicit but highly valuable for the comprehensibility of the process models.

The discovered process map allows to obtain operational knowledge about the structure of the process and the main patient-flows. Specifically, the analysis reveals that COVID-19 patients are characterized by a quite homogeneous high-level behavior, but several variants exist due to the possibility of a ICU admission or to the different outcomes of the process. More in detail, after the hospitalization and the onset of first symptoms, if present, each patient may be subject to both oxygen therapy and eventually ICU pathway, with subsequent ventilation and ECMO activities, until the end of the symptoms. Once conditions improve, patients may be discharged or transferred to another ward.

We evaluated the quality of the obtained process model through conformance checking~\cite{van2016data}. Specifically, we measured the token-based replay fitness between the Petri net and the event log, obtaining a value of 98\%. This is a strong indication of both a high level of compliance in the process (the flow of events does not deviate from the intended behavior) and a high reliability of the methodologies employed in data recording and extraction (very few deviations in the event log also imply very few missing events and a low amount of noise in the dataset).

\begin{figure}[h!]
\centering
\includegraphics[width=.75\textwidth]{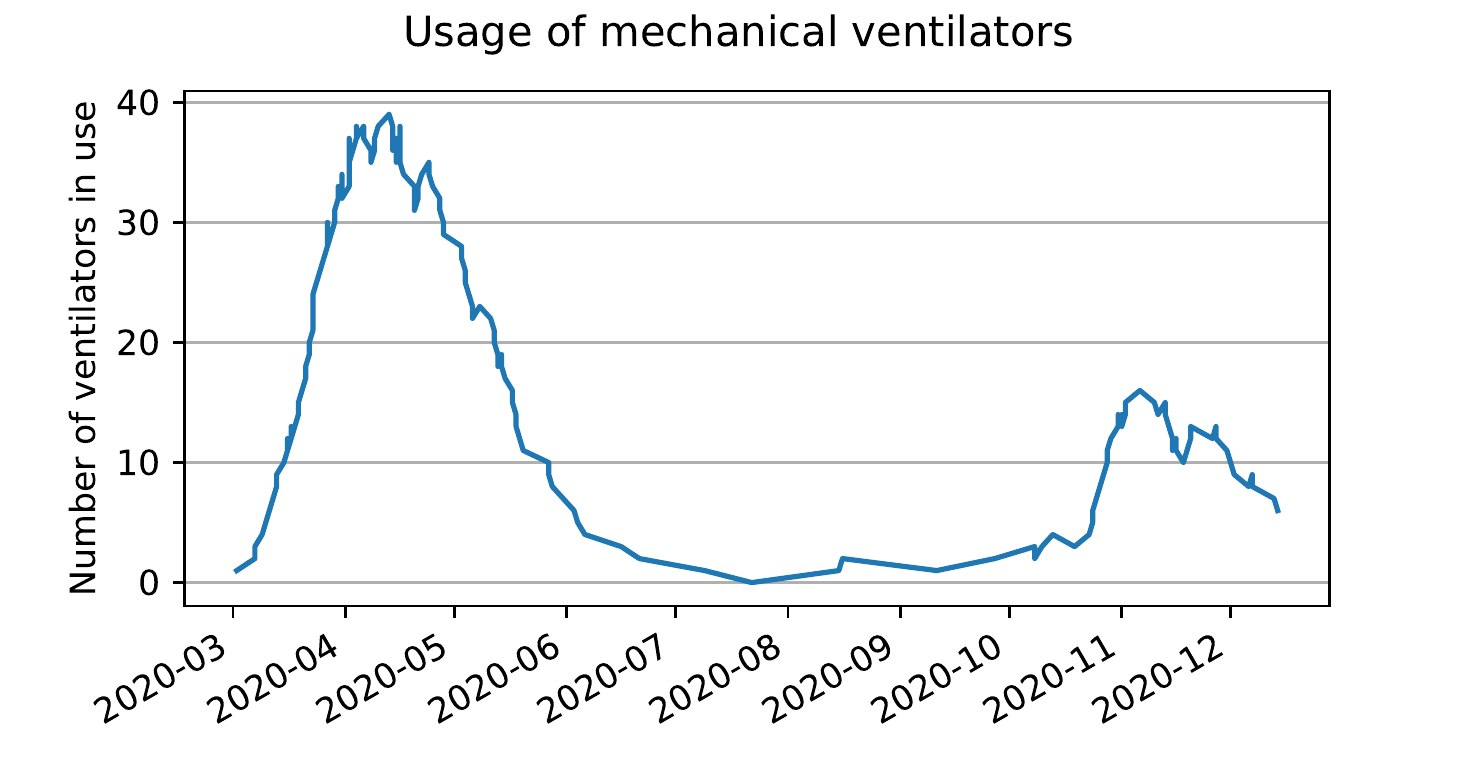}
\caption{Plot showing the usage of assisted ventilation machines for COVID-19 patients in the ICU ward of the Uniklinik Aachen. Maximum occupancy was reached on the 13th of April 2020, with 39 patients simultaneously ventilated.}
\label{fig:ventilation_covas}
\end{figure}

From the information stored in the event log, it is also possible to gain insights regarding the time performance of each activity and the resource consumption. For example, Figure~\ref{fig:ventilation_covas} shows the rate of utilization of ventilation machines.

\noindent This information may help hospital managers to manage and allocate resources, especially the critical or shared ones, more efficiently.

\begin{figure}[t]
	\centering
	\begin{minipage}[t]{0.48\textwidth}
		\centering
		\includegraphics[width=1\textwidth]{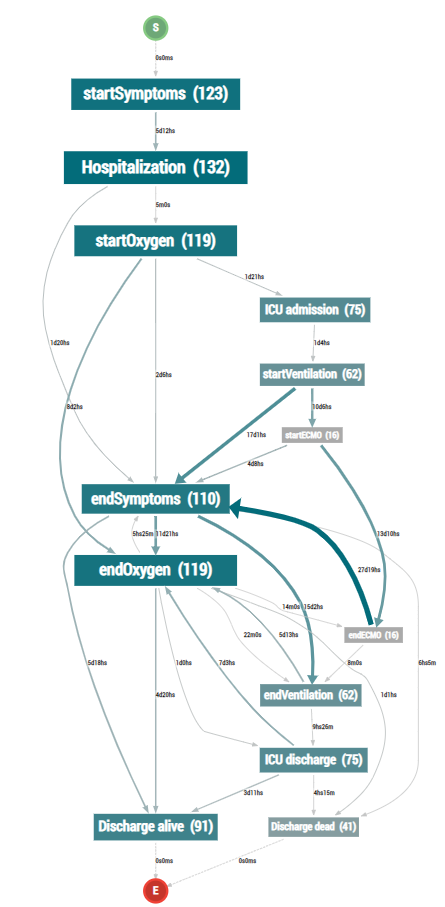}
		\caption{Filtered directly-follows graph related to the first wave of the COVID pandemic.}
		\label{fig:first_wave}
	\end{minipage}\hfill
	\begin{minipage}[t]{0.48\textwidth}
		\centering
		\includegraphics[width=1\textwidth]{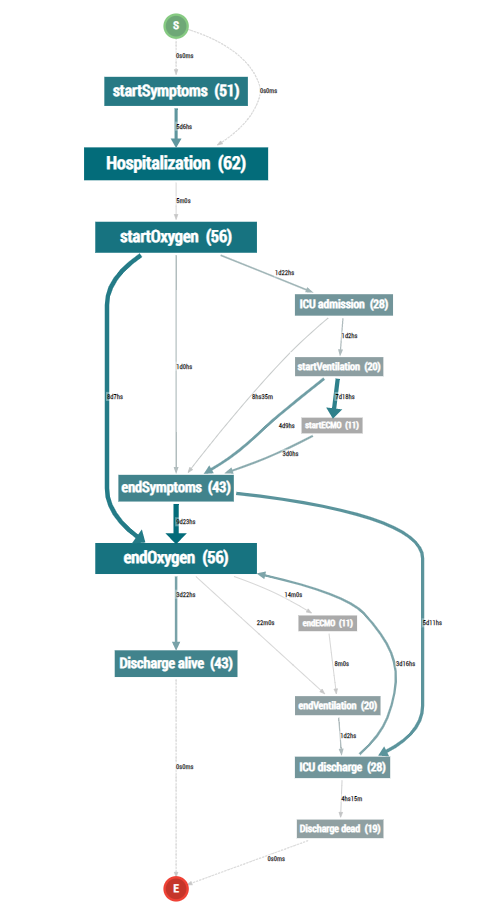}
		\caption{Filtered directly-follows graph related to the second wave of the COVID pandemic.}
		\label{fig:second_wave}
	\end{minipage}
\end{figure}

Finally, with the aid of the process mining tool Everflow~\cite{everflow}, we investigated different patient-flows, with respect to the first wave (until the end of June 2020) and second wave (from July 2020 onward) of the COVID-19 pandemic, and evaluated their performance perspective, which is shown in Figures~\ref{fig:first_wave} and~\ref{fig:second_wave} respectively. The first wave involves 133 cases with an average case duration of 33 days and 6 hours; the second wave includes 63 patients, with an average case duration of 23 days and 1 hour. The difference in average case duration is significant, and could have been due to the medics being more skilled and prepared in treating COVID cases, as well as a lower amount of simultaneous admission on average in the second wave.

\section{Conclusion and Future Work}\label{sec:conclusion}
In this preliminary paper, we show some techniques to inspect hospitalization event data related to the COVID-19 pandemic. The application of process mining to COVID event data appears to lead to insights related to the development of the disease, to the efficiency in managing the effects of the pandemic, and in the optimal usage of medical equipment in the treatment of COVID patients in critical conditions. We show a normative model obtained with the aid of IPD for the operations at the COVID unit of the Uniklinik Aachen hospital, showing a high reliability of the data recording methods in the ICU facilities.

Among the ongoing research on COVID event data, a prominent future development certainly consists in performing comparative analyses between datasets and event logs geographically and temporally diverse. By inspecting differences only detectable with process science techniques (e.g., deviations on the control-flow perspective), novel insights can be obtained on aspects of the pandemic such as spread, effectiveness of different crisis responses, and long-term impact on the population.

\bibliographystyle{splncs04}
\bibliography{bibliography}

\end{document}